# Adaptive Multicast of Multi-Layered Video: Rate-Based and Credit-Based Approaches[*]

Brett J. Vickers, Célio Albuquerque and Tatsuya Suda
{bvickers, celio, suda}@ics.uci.edu
Information and Computer Science Department
University of California, Irvine
Irvine, CA 92697-3425

## Abstract

*Network architectures that can efficiently transport high quality, multicast video are rapidly becoming a basic requirement of emerging multimedia applications. The main problem complicating multicast video transport is variation in network bandwidth constraints. An attractive solution to this problem is to use an adaptive, multi-layered video encoding mechanism. In this paper, we consider two such mechanisms for the support of video multicast; one is a rate-based mechanism that relies on explicit rate congestion feedback from the network, and the other is a credit-based mechanism that relies on hop-by-hop congestion feedback. The responsiveness, bandwidth utilization, scalability and fairness of the two mechanisms are evaluated through simulations. Results suggest that while the two mechanisms exhibit performance trade-offs, both are capable of providing a high quality video service in the presence of varying bandwidth constraints.*

## 1 Introduction

In an era of proliferating multimedia applications, support for video transmission is rapidly becoming a basic requirement of network architectures. It has long been recognized that high speed networking technologies like ATM are capable of supporting the strict quality of service guarantees required by real-time traffic like video. Yet even in networks that have traditionally offered minimal or no quality of service guarantees, efforts are now underway to support real-time video applications. Quality of service support in the Internet, for instance, is the subject of a great deal of recent research attention [1].

Furthermore, since most video applications (e.g., teleconferencing, television broadcast, video surveillance) are inherently multicast in nature, support for point-to-point video communication is not sufficient. Unfortunately, multicast video transport is severely complicated by variation in the amount of bandwidth available throughout the network. See the example shown in Figure 1. The video source V attempts to transmit video to two destinations, $D_1$ and $D_2$, at a peak rate of 20 Mbps, but due to competing network traffic and varying link capacities, the path between V and $D_1$ can support 10 Mbps of video, while the path between V and $D_2$ can support only 4 Mbps. One potential solution to this problem of varying bandwidth constraints is to force the source to apply an adaptive video encod-ing technique and reduce its transmission rate to 4 Mbps, which is the highest rate that both paths can support. However, in a multicast connection with hundreds or even thousands of destinations, there is likely to be at least one very congested path. Limiting the video rate according to the most congested path penalizes the quality of video offered across all the other paths, regardless of how much bandwidth is available on them.

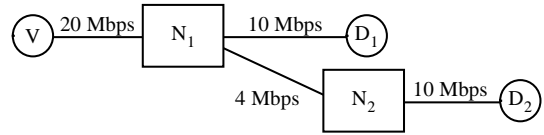

Figure 1: Example multicast video session.

A more scalable solution to the problem of available bandwidth variation is to use multi-layered video. A multi-layered video encoder encodes raw video data into one or more streams, or layers, of differing priority. The layer with the highest priority, called the base layer, contains the most important portions of the video stream. One or more enhancement layers with progressively lower priorities may then be encoded to further refine the quality of the base layer stream. For instance, in the example of Figure 1, the ideal deployment of multi-layered video results in a base layer stream transmitted at 4 Mbps and a single enhancement layer stream transmitted at 6 Mbps.

There are two primary advantages to using multi-layered video encoding in multicast-capable networks. First is the ability to perform graceful degradation of video quality when loss occurs. Because each video layer is prioritized, a network experiencing congestion may discard packets from low priority layers, thereby protecting the important base layer and higher priority enhancement layers from corruption. The second advantage, which is related to the first, is the ability to support multiple destinations with different bandwidth constraints or end-system capabilities. For each source-to-destination path with a unique bandwidth constraint, an enhancement layer of video may be generated.

Multi-layered video is not by itself sufficient to provide ideal network bandwidth utilization or video quality, however. To improve the bandwidth utilization of the network and optimize the quality of video received by each of the destinations, the source must respond to constantly changing network conditions by dynamically adjusting the number of video layers it generates as well as the rate at which each layer is transmitted. For the source to do this, it must have congestion feedback from the destinations and the network.

In this paper, we study two novel and promising feedback mechanisms, both of which rely on adaptive, multi-layered video encoding. The first is a rate-based mechanism that uses a closed feedback loop, where the source periodically multicasts feedback packets to each destination, and the destinations return the packets to the source. As these feedback packets traverse the network, intermediate nodes examine their current state of congestion and determine the number of layers the video source should generate as well as the explicit rates of each layer. To prevent feedback implosion, intermediate nodes merge returning feedback packets. When the source receives a returning feedback packet, it adjusts its encoding behavior to generate the specified number of layers at the specified rates.

---

[*] This research is supported by the National Science Foundation through grant NCR-9628109. It has also been supported by grants from the University of California MICRO program, Hitachi Ltd., Hitachi America, Standard Microsystem Corp, Tokyo Electric Power Company, Nippon Telegraph and Telephone Corporation (NTT), Nippon Steel Information and Communication Systems Inc. (ENICOM), Matsushita Electric Industrial Company, and Fundação CAPES/Brazil.

The second feedback mechanism is a credit-based mechanism that uses hop-by-hop flow control to reduce loss and optimize utilization. Intermediate nodes exchange feedback packets containing "credits," which reflect the amount of buffer space available at the next downstream node. Feedback packets also propagate congestion control information from the destinations to the source. By the time the source receives a feedback packet, it is aware of exactly how many destinations are fully or partially receiving each video layer. It uses this information to adjust the number of video layers it generates as well as the transmission rate of each video layer.

The remainder of this paper is organized as follows. Related research on the transport of video traffic in high speed networks is reviewed in section 2. The multicast, multi-layered feedback mechanisms introduced by this paper are detailed in Section 3. The responsiveness, utilization, scalability, and fairness of the two feedback mechanisms are evaluated through simulations in section 4, and concluding remarks are provided in section 5.

## 2 Related Work

A number of researchers have examined the use of congestion feedback for the adaptive control of the video encoding process [2-5]. In [2], [3] and [4] information regarding the occupancies of internal network buffers is passed via network feedback packets to the video source. The encoding of the video sequence is then rate-controlled to avoid buffer overflow within the network. In [5], network switches implement an explicit rate control policy and inform the video source of the exact rate at which to encode video, thereby rapidly adjusting to changes in the network's available bandwidth due to transient congestion effects. However, in none of these works is the specific problem of transmitting multicast video across paths with varying bandwidth constraints taken into account.

In another work [6], a scenario in which a single end system transmits a single layer of video to several IP destinations is considered, and congestion feedback from the destinations is used to control the rate of the video stream. A form of probabilistic feedback used to prevent feedback implosion. Based on feedback responses from the destinations, the source adaptively modifies the video encoding rate to reduce network congestion when necessary and increase video quality where possible. While this scheme takes multicast connections into account, it uses only a single layer of video, and thus a few severely bandwidth-constrained paths can negatively impact the rate of video transmitted across paths that have more plentiful bandwidth.

The destination set grouping approach [7] attempts to satisfy the bandwidth constraints of multiple source-to-destination paths in the distribution of multicast video. The source maintains a small number of independent video streams, each encoded from the same raw video material but at different rates. The video streams are then targeted to destination groups with different bandwidth constraints. Feedback from the destinations is used to control the encoding rates of each offered video stream, and destinations are allowed to choose which stream to receive based on their current bandwidth constraints. Although this multicast approach is adaptive, transmitting several independently encoded video streams may result in an inefficient use of network bandwidth.

Another potential solution to the multicast of video to destinations with varying bandwidth constraints is transcoding [8]. In this approach, a single layer of video is encoded at a high rate by the source, and intermediate network nodes transcode (i.e., decode and re-encode) the video down to a lower rate whenever they become bottlenecked. While this approach solves the available bandwidth variation problem, it requires complex and computationally expensive video transcoders to be present throughout the network.

The receiver-driven layered multicast (RLM) approach for IP networks [9] is perhaps the one most closely related to this paper's proposed mechanisms. In the RLM approach, the source generates a fixed number of layers, each at a fixed rate, and the destinations "subscribe" to as many layers as they have the bandwidth to receive. This approach, while it improves the efficiency of video transport through multi-layered encoding, is not adaptive; the destinations choose among the layers the source is willing to provide. Unfortunately, in some cases the provided selection may not be adequate enough to optimize network utilization and video quality.

The adaptive approaches described in this paper use feedback from the network to optimize both the network utilization and the quality of video received by the destinations. This work is a significant extension of the authors' prior work [10], in which a rate-based, two-layer video encoding technique was applied. In that work, congestion indications were based on network buffer occupancies and were binary in nature. This paper presents two novel feedback mechanisms — an explicit rate-based approach and a credit-based approach — both of which allow for an arbitrary number of encoded video layers.

## 3 Mechanisms

The rate-based and credit-based mechanisms for the multicast of adaptively encoded video are described in detail below.

### 3.1 Rate-Based Mechanism

To satisfy a large number of video multicast destinations with varying bandwidth constraints, a rate-based, closed loop congestion control algorithm is introduced. Through the exchange of congestion feedback with the network, the video source learns how many video layers to generate as well as the transmission rates for each layer.

In a closed loop feedback algorithm like the one being proposed, the source periodically generates a control packet called a "forward feedback packet," which it sends to the destinations. Upon receiving the forward feedback packet, a destination copies the packet's contents into a "backward feedback packet" and returns it to the source, thereby closing the feedback loop. To maintain a steady flow of feedback between the source and the destination, the source generates one new forward feedback packet for every $N_f$ video packets sent, where $N_f$

| Field | Description | Used in forward feedback packets | Used in backward feedback packets |
|---|---|---|---|
| $L$ | Maximum number of video layers allowed | √ | √ |
| $R_C$ | Current combined rate of the video source | √ | |
| $R_E$ | The maximum explicit rate allowed on the path | √ | |
| $N_t$ | Current number of video layers | | √ |
| $r_i$ | An array ($i = 1, ..., N_t$) listing the cumulative rates of each video layer | | √ |
| $c_i$ | An array ($i = 1, ..., N_t$) listing the number of destinations requesting each layer in the rate array $r_i$ | | √ |

Table 1: Contents of feedback packets used by the rate-based mechanism

is a relatively small number such as 16.

As feedback packets traverse the closed loop, intermediate network nodes mark them in order to explicitly indicate the amount of bandwidth available in the network for the transmission of video. The intermediate nodes must therefore (1) monitor the amount of bandwidth available for video, (2) track the number of video multicast connections attempting to share the available bandwidth, and (3) calculate the fair share of the available bandwidth for each video multicast connection competing for the outgoing link. An existing algorithm, known as the Explicit Rate Indication for Congestion Avoidance (ERICA) algorithm [11], has been devised to support these functions in ATM networks for Available Bit Rate (ABR) data services, and we adopt it as part of the proposed rate-based, multicast feedback mechanism. Most of ERICA's functions take place in intermediate network nodes, where the available bandwidth is monitored and feedback packets are marked. The functions that ERICA performs in the output ports of intermediate nodes are briefly summarized as follows:

1. Set the target utilization of the link bandwidth to some fraction of the total link capacity (e.g., 95%). A target utilization less than 100% helps the switch prevent buffer overflows due to transient congestion effects. It also shortens queueing delays by keeping buffer occupancy low.

2. Monitor the number of active ABR virtual connections.

3. Monitor the amount of non-ABR guaranteed traffic arriving at the output port and calculate the amount of bandwidth remaining for use by ABR traffic. This amount is known as the "ABR capacity."

4. Monitor the amount of ABR traffic arriving at the port's output queue, and calculate the "overload." The overload is equal to the ABR input rate divided by the ABR capacity and measures the degree to which ABR traffic is congesting the link.

5. Using the overload value, calculate the "VC share," which is equal to the virtual connection's current cell rate divided by the overload. The VC share represents an allocation of bandwidth that restores the link to the target utilization. It optimizes utilization of the link during periods of underload and prevents loss during periods of overload.

6. Calculate the connection's "fair share" of the available bandwidth. The fair share is equal to the ABR capacity divided by the number of active ABR connections.

7. Set the explicit rate ($R_E$) value for the connection to the larger of the "VC share" and the "fair share." Place the explicit rate value into the forward feedback packet, but do not allow it to exceed the ABR capacity of the link or the explicit rate value calculated by the previous hop.

All of ERICA's monitoring operations take place over the duration of a short, fixed averaging interval. New values for the overload, fair share, VC share and explicit rate are computed only once per averaging interval.

The two primary goals of the ERICA algorithm are to optimize bandwidth utilization and the fairness of the bandwidth allocation. We believe that ERICA, while originally devised for bursty data traffic in ATM networks, is also applicable in more general circumstances. More specifically, it is applicable to the multicast of adaptively encoded video, for which video quality depends on the fair and efficient utilization of network bandwidth. We have therefore applied ERICA to the congestion and flow control of multi-layered, multicast video.

Table 1 lists the fields contained within each of the proposed rate-based mechanism's feedback packets. When a forward feedback packet is generated, the source stores the maximum number of video layers it can support ($L$). The value of $L$ depends on the the number of layers the video encoder is able to generate. For example, if the source uses a scalable encoder that can only generate four layers of video (one base layer plus three enhancement layers), then it sets $L$ to 4. The value of $L$ must also be less than or equal to the maximum number of priority levels the network can support. The current combined rate ($R_C$) field contains the combined rate of all video layers currently being generated by the source. This field is used by the ERICA algorithm to calculate the VC share. Finally, the explicit rate field ($R_E$) is set by intermediate nodes according to the ERICA algorithm and indicates to each of the destinations how much bandwidth is available on the path from the source. At feedback packet generation time, the source initializes the explicit rate to the peak rate of the connection.

As forward feedback packets pass through intermediate nodes on the way to their destinations, they are copied to multiple output links, just as video packets are. The intermediate nodes monitor the amount of bandwidth available for video on each outgoing link and use the ERICA algorithm to divide that bandwidth fairly between all competing multicast video connections. After an intermediate node determines the amount of bandwidth to allocate to the connection, it enters the value into the feedback packet's explicit rate ($R_E$) field. This process is repeated at each of the subsequent intermediate nodes.

Upon receiving a forward feedback packet, the destination examines the explicit rate field to determine how much bandwidth is available for video. Since the available bandwidth varies from branch to branch of the multicast connection, each destination is likely to see a different explicit rate value. The destination then generates a "backward feedback packet" and sets its contents to indicate the desired video rate. It does this by filling the first slot of the backward feedback packet's rate array ($r_1$) with the explicit rate value contained in the forward feedback packet. It also sets the corresponding slot's counter ($c_1$) to one in order to indicate that only one destination has requested rate $r_1$ so far.

Closed loop feedback mechanisms cause destinations to return one backward feedback packet for every forward feedback packet received. In a multicast environment, this can result in *feedback implosion*, where the source receives a large number of bandwidth-consuming feedback packets. To prevent implosion, the junction intermediate nodes *merge* backward feedback packets as they return from destinations.[1] A merging operation is relatively simple to implement; it merely requires the intermediate node to wait for one of the following two conditions to become true: (1) at least one backward feedback packet has arrived from each of the connection's downstream nodes, or (2) a feedback merging timer for the connection has expired. Once one of these conditions is fulfilled, the junction node collects the rate ($r_i$) and counter ($c_i$) entries from each feedback packet and stores them into a temporary local array, sorted by rate. Each rate entry corresponds to a video rate requested by one or more downstream destinations, while the counter values indicate how many downstream destinations have requested each rate. Ultimately, the rate values will be used by the source to determine the rates to transmit each video layer. If two or more packets contain identical rate values (or nearly identical values[2]), then their corresponding counter values are summed together and stored with the rate as a single local array entry.

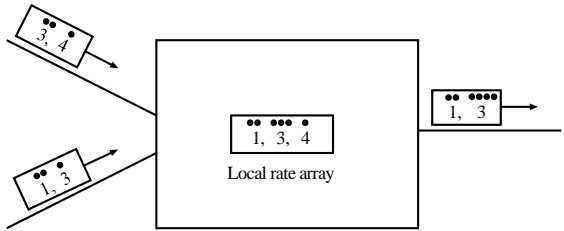

Figure 2: Example of backward feedback packet merging ($L=2$)

After filling the local rate array, the number of entries in the array is compared to the maximum number of layers allowed for the connection ($L$). If the number of entries in the local rate array is less than or equal to the maximum number of layers allowed, then a new backward feedback packet is immediately generated, filled with the contents of the local rate array, and sent to the next hop. Otherwise, one (or more) of the entries must be discarded and its counter values added to the next lower entry. To determine which entry (or entries) to discard, the intermediate node attempts to estimate the impact of dropping each listed rate on the overall video quality. This is done through the use of a simple estimated video quality metric.

The estimated video quality metric attempts to measure the combined "goodput" of video traffic that will be received by all downstream destinations. The goodput for a single destination is defined as the total throughput of all video layers received by the destination *without loss*. For instance, suppose a source transmits three layers of video at 1 Mbps each. If a destination entirely receives the most important first two layers but only receives half of the third layer due to congestion, then its total received throughput is 2.5 Mbps, but its *goodput* is equal to the combined rate of the first two layers, namely 2 Mbps. The goodput is a relatively useful estimate of video quality because it measures the total combined rate of uncorrupted video traffic arriving at an end system.

As intermediate nodes merge backward feedback packets, they attempt to estimate the goodput that downstream destinations will receive. The combined goodput $G$ is estimated from the values listed in the rate array and is summed over $N$ as follows: $G = \sum r_i \times c_i$, where $N$ is the number of entries in the local rate array, and $r_i$ and $c_i$ are the rate and counter values for each entry. To determine which entry to remove from the local rate array, it is necessary to calculate the combined goodput that will result from each potential entry removal. The entry removal that results in the highest combined goodput is then removed from the rate array. This process is repeated until the number of entries in the local rate array is equal to the maximum number of layers allowed. The number of entries in the rate array ($N_l$) is set to $L$, and a merged feedback packet is transmitted to the next hop.

(There is one important caveat when removing an entry from the local rate array: the first entry can never be removed. Even minor losses in the base layer can cause precipitous drops in video quality, so the base layer should ultimately reflect the amount of bandwidth available on the most congested path. Hence, the array entry with the lowest rate can never be removed, because it may ultimately determine the rate of the base layer.)

For an example of the feedback merging process used by the rate-based mechanism, consider Figure 2. Two backward feedback packets are shown arriving at an intermediate node, both with two rate entries ($r_1$ and $r_2$) in units of Mbps stored in their rate arrays. The counter values ($c_i$) are indicated by the number of dots over each listed rate. Since both packets contain a rate entry of 3 Mbps, these entries are merged into a single entry in the local array, and their counter values of 1 and 2 are added together, as shown, in order to indicate that three downstream destinations have requested a rate of 3 Mbps. After storing the feedback packets' entries into the local rate array, one entry must be removed to bring the total number of rates down to 2, which is the maximum number of layers allowed for this example. Since the first entry can never be removed, this leaves only the second and third entries as candidates for removal. If the second entry is removed, then its counter value will be added to the first entry and the resulting combined goodput $G$ will be $(1 \times 5) + (4 \times 1) = 9$. If the third entry is removed, then its counter value will be added to the second entry, and the resulting goodput will be $G = (1 \times 2) + (3 \times 4) = 14$. Since the removal of the third entry results in a higher combined goodput than the removal of the second entry, the third entry is removed. The resulting backward feedback packet contains two rate entries and is forwarded to the next hop.

By the time a feedback packet arrives at the source, it contains the number of video layers to encode and a list of cumulative rates at which to encode each layer. This completes one cycle of the feedback loop.

The effect of this rate-based feedback mechanism is to dynamically establish the number of video layers to encode as well as nearly optimal rates for each of the layers. The rates are optimal in the sense that they are selected by the network in a manner that optimizes the combined goodput. Under the rate-based mechanism, bandwidth in the network is almost fully utilized, and the quality of video received by most of the destinations is determined not solely by the source or the receiver, but also by the current state of congestion in the network.

---

[1] In this paper, it is assumed that the network is capable of establishing a multicast connection and that feedback packets traveling from the source to each destination traverse the same intermediate nodes on their return paths.

[2] Two rate values that are separated by less than 100 kbps are considered the same rate, and the lesser of the two rates is stored in the local rate array.

| Field | Description |
|---|---|
| $L$ | Maximum number of video layers allowed |
| $C$ | Credit counter, indicating the number of credits sent so far to the upstream node |
| $F_i$ | An array ($i = 1, ..., L$) listing the number of downstream destinations fully receiving up to video layer $i$ and partially receiving zero layers |
| $P_i$ | An array ($i = 1, ..., L$) listing the number of downstream destinations fully receiving up to video layer $i$ - 1 and partially receiving layer $i$ |

Table 2: Contents 0f feedback packets used by the credit-based mechanism

## 3.2 Credit-Based Mechanism

In addition to the rate-based mechanism, we also investigate a credit-based mechanism for the adaptive multicast of multi-layered video. Credit-based mechanisms have been widely studied, especially in regard to the flow and congestion control of data traffic [12,13]. The credit-based scheme proposed in this paper for multi-layered video is influenced largely by the *Quantum Flow Control* (QFC) mechanism [12] used for ABR data traffic in ATM networks. The primary advantage of QFC is its ability to achieve 100% network utilization while ensuring zero packet loss, regardless of the amount of network congestion.

The QFC mechanism maintains a separate control loop for each link of a connection by using *credits*. Credits reflect the amount of buffer space available at the next downstream node and give a node permission to transmit packets. Each time a node transmits a packet, it consumes one credit and transfers a credit to the connection's upstream node. If a node has no credits available, it must wait for one or more credits to arrive before transmitting a packet. To prevent the inefficient use of bandwidth by credit packets, several credits are collected by each node before being transmitted together to an upstream node.

For multicast connections, the original QFC algorithm is designed to reduce the source's transmission rate in response to the connection's most congested branch. For multi-layered video, this type of behavior is undesirable since full utilization of network bandwidth is one of the primary goals, and losses to low priority video layers are tolerable. This paper introduces a modified credit-based mechanism that extends QFC to achieve full utilization on all branches of a multicast connection. In the modified credit-based mechanism, losses are allowed to occur, but when buffers overflow, only the packets from the lowest priority layers are discarded. Destinations also supply feedback in order to allow the source to determine which destinations are fully or partially receiving each layer and thereby adjust the number of layers as well as the rate of each layer. A detailed description of the credit-based mechanism for multicast video follows.

An intermediate node returns a feedback packet to its upstream neighbor whenever one of the following two conditions is satisfied:

1. Each of the multicast connection's output ports has transmitted at least $N_t$ packets, or
2. At least one output port of a multicast connection has transmitted $N_t$ packets, and the difference between the occupancies of any two video output queues in the same multicast connection is at least $D_t$ packets.

The first condition guarantees that credits are periodically returned to an upstream node whenever each of the connection's adjacent downstream nodes is continually draining packets. The second condition allows credits to be returned to an upstream node even if one or more adjacent downstream nodes fails to drain packets rapidly enough. This second condition prevents a node from becoming a bottleneck as long as at least one downstream path continues to accept packets. While this condition may result in packet losses on some links, the losses are isolated to low priority packets through a priority discard mechanism. In both conditions, feedback packets carry $N_t$ credits to the node's upstream neighbor, which increments its credit counter by $N_t$.

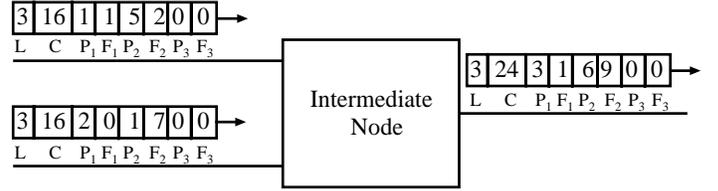

Figure 3: Feedback packet generation at intermediate nodes.

Table 2 lists the information carried by each of the proposed credit-based mechanism's feedback packets. $C$ is equal to the total number of credits that the downstream node has sent to the upstream node since call establishment, and $L$ is the maximum number of video layers that can be generated by the source and transported by the network. The feedback packet also contains two arrays of counters, the full reception array ($F_i$) and the partial reception array ($P_i$), which are ultimately used to indicate to the source the number of destinations that are fully and partially receiving each video layer. Within a given time interval, a layer is said to be "fully received" if its packets arrive at the destination totally uncorrupted by packet loss, whereas a layer is considered "partially received" if over 25% of that layer's packets are received.[3] If fewer than 25% of a layer's packets are received by the destination, then the layer is considered neither fully nor partially received. Each $F_i$ entry indicates the number of downstream destinations that are fully receiving all layers up to layer $i$ and are not partially receiving any layers. Each $P_i$ entry indicates the number of downstream destinations that are fully receiving up to layer $i$ - 1 and partially receiving layer $i$. For example, a value of $F_3$ equal to 2 indicates that two downstream destinations have not partially received any layers and have fully received layers 1, 2 and 3. A value of $P_3$ equal to 1 indicates that one downstream destination has fully received layers 1 and 2 and has only partially received layer 3.

Every time a destination receives $N_t$ video packets, it generates a feedback packet containing $N_t$ credits and sets an entry in one of the reception arrays ($F_i$, $P_i$) to 1. In order to determine which reception entry to set, the destination monitors the arrival of video packets over a sliding interval of time called the *reception monitoring interval*, which is long enough to enable the reception of packets from all layers. If any layer is partially received during the monitoring interval, the $P_i$ entry for that layer is set. If no layers are partially received, then the destination sets an $F_i$ entry for the lowest priority layer fully received.

When an intermediate node generates a feedback packet, it computes new $F_i$ and $P_i$ entries for the packet. For each layer $i$, it stores the sum of the $F_i$ entries from the connection's arriving feedback packets into the new packet's $F_i$ entry. The same operation is performed for the $P_i$ entries. Hence, intermediate nodes accumulate the number of downstream destinations that have fully and partially received each layer $i$. Figure 3 shows an example.

---

[3] While a value of 25% was used in this paper, another value may be used to determine partial reception.

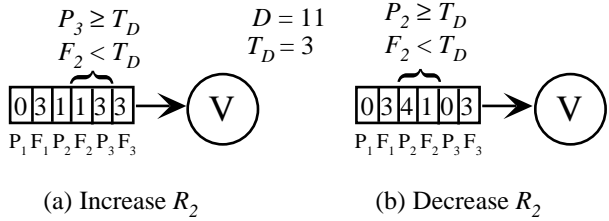

(a) Increase $R_2$    (b) Decrease $R_2$

Figure 4: Example application of the second set of rules for adjusting layer rates ($L=3$)

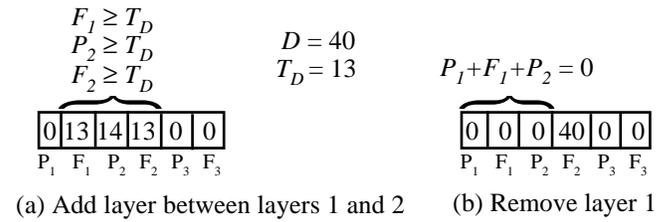

(a) Add layer between layers 1 and 2    (b) Remove layer 1

Figure 5: Example application of the third set of rules for creating and deleting layers ($L=3$, four feedback packets collected)

By the time the source receives a feedback packet, it is aware of exactly how many destinations are fully and partially receiving each video layer. The source then uses this information, as well as its buffer occupancy, to adjust the number of video layers and the transmission rate of each video layer. It does this by applying several simple rules, which are divided into three sets. The first set of rules controls the rate of the lowest priority video layer, the second set of rules controls the rate of each remaining video layer, and the third set of rules controls the creation and deletion of video layers.

The first set of rules depends only on the current occupancy of the source buffer and is applied whenever a feedback packet arrives. This set of rules controls the rate of the lowest priority layer and, thereby, the overall video transmission rate by monitoring the occupancy of the source buffer, which has three thresholds. The lower threshold helps the source detect when it is generating video at a bit rate lower than the network can accept, the middle threshold serves as a target occupancy for the source buffer, and the upper threshold is used by the source to detect when it is generating video at a rate higher than the network can accept. The following rules attempt to hold the source buffer occupancy between the lower and middle thresholds by adjusting the rate of the lowest priority video layer:

1. If the source buffer occupancy is less than the lower threshold, the rate of the lowest priority video layer is incremented by a constant value in order to prevent buffer underflow.

2. Conversely, if the source buffer occupancy exceeds the upper threshold, then the source attempts to rapidly reduce the rate of its lowest priority video layer in order to avoid buffer overflow. First, it calculates the rate $R_N$ at which the first hop is draining packets:

$$R_N = \frac{N_t \times \text{Packet\_size}}{t - t_{pf}}$$

where $t$ is the current time and $t_{pf}$ is the arrival time of the previous feedback packet. If $R_N$ is less than the current overall video transmission rate, then the rate of the lowest priority video layer is adjusted so that the current overall rate equals $R_N$. Otherwise the rate of the lowest priority video layer is decremented.

3. If the buffer occupancy is between the upper and middle thresholds, then the rate of the lowest priority video layer is decremented in an attempt to bring the buffer occupancy down to the region between the lower and middle thresholds.

4. If the buffer occupancy falls between the lower and middle thresholds, and the buffer occupancy is higher than it was at time $t_{pf}$, then the rate of the lowest priority video layer is decremented by a constant value. Otherwise the rate of the lowest priority video layer is incremented. This rule attempts to keep the buffer occupancy between the lower and middle thresholds as much as possible.

Please note that this is not the only set of rules that could have been applied to the source buffer. Simpler two- or one-threshold schemes may also be used.

The first set of rules only changes the rate of the lowest priority video layer. In order to adjust the rates of the remaining layers, a second set of rules is applied whenever feedback packets arrive at the source. Based on the values contained in the $F_i$ and $P_i$ arrays, the source classifies destinations into groups, with one group defined for each non-zero reception array entry. The source uses a threshold $T_D$ to decide when too many destinations are partially receiving a given layer. $T_D$ is computed as follows: $T_D = \lfloor D / L \rfloor$, where $L$ is the maximum number of video layers, and $D$ is the number of destinations that participated in filling the contents of the feedback packet. ($D$ is equal to the sum all the entries in the $P_i$ and $F_i$ arrays.) The cumulative encoding rate $R_i$ for each layer $i$ is then adjusted according to the following rules:

1. If $P_i \geq T_D$ and $F_{i-1} < T_D$, then the cumulative rate of layer $i-1$ is incremented. The example in Figure 4(a) illustrates this rule. In this example, three destinations are partially receiving layer 3 ($P_3 = 3$), while only one destination is fully receiving layer 2 ($F_2 = 1$). This indicates that more destinations would benefit from a *higher* cumulative rate for layer 2 than would be harmed by it. The cumulative rate of layer 2 is therefore increased so that more destinations may fully receive it.

2. Otherwise, if $P_i \geq T_D$ and $F_i < T_D$, then the cumulative rate of layer $i$ is decremented. The example in Figure 4(b) shows four destinations partially receiving layer 2 ($P_2 = 4$) and only one destination fully receiving layer 2 ($F_2 = 1$). This indicates that more destinations would benefit from a *lower* cumulative rate for layer 2 than would be harmed by it. The cumulative rate of layer 2 is therefore decreased so that more destinations may fully receive it.

In some cases, video layers may also need to be created or deleted. Over the period of a fixed time interval called the *feedback accumulation interval*, the source accumulates totals for each of the full and partial reception array entries ($F_i$ and $P_i$). At the start of each new interval it determines whether to create or delete video layers. By accumulating the information received in feedback packets, the source avoids creating or deleting layers in response to transient congestion fluctuations in the network.

The source adjusts the number of layers by using the third set of rules, described below:

1. If $F_{i-1} \geq T_D$ and $P_i \geq T_D$ and $F_i \geq T_D$, then a new layer is created at a cumulative rate between $R_{i-1}$ and $R_i$. Figure 5(a) illustrates this rule through an example in which the contents of four feedback packets have been accumulated during a single feedback accumulation interval. The source sees that layers 1 and 2 have been fully received by an accumulated total of 13 destinations ($F_1 = F_2 = 13$), and that layer 2 has been partially received by an accumulated total of 14 destinations ($P_2 = 14$). The fact that all three values are greater than or equal to the accumulated value of $T_D$ ($\lfloor 40/3 \rfloor = 13$) indicates that a disproportionate number of destinations are being served by layers 1 and 2. Hence, the source creates a new layer with a cumulative rate between $R_1$ and $R_2$.

2. If $P_i + F_i + P_{i+1} = 0$, then layer $i$ is not serving any destination and is consequently removed. The example of Figure 5(b) shows an accumulated total of 40 destinations fully receiving layer 2 ($F_2 = 40$), no destinations partially receiving layer 2 ($P_2 = 0$), and no destinations being served by layer 1 ($F_1 = P_1 = 0$). Layer 1 may therefore be removed without affecting the video quality delivered to any destination.

As in the rate-based feedback mechanism, the credit-based mechanism dynamically adjusts the number of video layers as well as the rates of each of the layers. Video quality and bandwidth utilization are determined by the source, the network and the destinations.

## 4 Performance

This section presents the results of several simulations designed to evaluate the performance of the two proposed multicast, multi-layered feedback mechanisms. We use several network topologies to evaluate performance metrics including the responsiveness, utilization, scalability and fairness of the rate-based and credit-based mechanisms. All simulations assume the use of ATM cell-sized packets. Unless otherwise specified, all link capacities are equal to 100 Mbps, propagation delays between end systems and intermediate nodes are 5 μs, and propagation delays between intermediate nodes are 5 ms.

For the rate-based mechanism, a target utilization of 99% is assumed, and forward feedback packets are transmitted once for every 15 video packets transmitted ($N_f = 15$). Intermediate nodes use an ERICA averaging interval of 10 ms and a feedback merging time-out interval of 50 ms. Unless otherwise specified, buffer sizes of 200 packets per multicast connection are allocated.

For the credit-based mechanism, feedback packets are generated once for every 8 packets transmitted or when the difference between the occupancies of any two video buffers for the same multicast connection is 8 ($N_t = D_t = 8$). When layer transmission rates are incremented or decremented, the size of the adjustment is 16 packets per second. A reception monitoring interval of 10 ms is used at each of the destinations, and a feedback accumulation interval of 40 ms is used by the source. The size of the source buffer is 600 packets, with thresholds at 20, 200 and 300, and intermediate node buffers of 300 packets per multicast connection are allocated.

### 4.1 Responsiveness

In order to be effective, feedback-based traffic control mechanisms must react in a timely fashion to changes in the network's congestion status. Both the rate-based and credit-based mechanisms attempt to react rapidly to changes in the network's available bandwidth by adjusting the number of video layers the source generates as well as the rate of each layer.

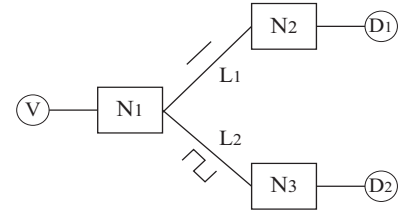

Figure 6: Simulation model for evaluating responsiveness.

A tree topology network model is used to evaluate responsiveness. As shown in Figure 6, it consists of one source $V$, two destinations $D_1$ and $D_2$, and three intermediate nodes $N_1,...,N_3$. Interfering traffic is applied on the links connecting intermediate nodes, and two responsiveness experiments are conducted. The first experiment is organized so that the source is required to create and delete video layers in response to changes in the available bandwidth in the network. The second experiment is designed to require the source to adjust the rate of one of its video layers in response to changes in network congestion.

In the first experiment, a persistent stream of constant rate interfering traffic is applied to link $L_1$. The transmission rate of this interfering stream is 90 Mbps, leaving 10 Mbps of available bandwidth for use by video traffic. On link $L_2$, square-wave interfering traffic that oscillates every two seconds between constant rates of 90 and 95 Mbps is applied in order to test the responsiveness of the source to rapid changes in the network's available bandwidth.

Figure 7 displays the rates of video traffic layers generated by the source for the rate-based and credit-based mechanisms. For the first two seconds of the simulation, 10 Mbps is available for video on both bottleneck links. In the case of the rate-based mechanism, this results in a single layer of video transmitted at 9 Mbps. Since the rate-based mechanism relies on explicit rate notifications, it takes approximately one round trip time of 10 ms to converge to 9 Mbps. The reason the full 10 Mbps of available bandwidth is not used by the rate-based source is because the ERICA algorithm employed in this simulation establishes a target utilization of 99%, and hence 1 Mbps of the link's bandwidth remains unutilized. At time $t = 2$ sec, the available bandwidth on link $L_2$ drops from 10 Mbps to 5 Mbps, and the rate-based mechanism takes one round trip time to react. The source reduces the rate of the base layer to 4 Mbps and generates a new enhancement layer at a cumulative rate of 9 Mbps. At time $t = 4$ sec, the available bandwidth on link $L_2$ returns to 10 Mbps, and the enhancement layer is removed 10 ms later. As the available bandwidth on link $L_2$ oscillates between 5 and 10 Mbps, the rate-based mechanism responds by cyclically adding and removing a video layer within a single round trip time, as shown in Figure 7(a).

The credit-based mechanism responds somewhat more slowly to changes in available bandwidth in the network. As Figure 7(b) shows, the credit-based source takes approximately 1.8 seconds to converge to the available bandwidth of 10 Mbps at the start of the simulation. When the available bandwidth on link $L_2$ drops to 5 Mbps at time $t = 2$, the credit-based mechanism responds 80 ms later by generating two layers. It then takes an additional 10 ms for the enhancement layer to reach a cumulative rate of 10 Mbps and an additional 200 ms for the base layer to drop to 5 Mbps. At time $t = 4$ sec, when the available bandwidth on link $L_2$ returns to 10 Mbps, the credit-based mechanism allows the source to respond relatively quickly. It removes the enhancement layer within 40 ms of the change in network congestion and immediately transmits a single layered stream at a rate of 10 Mbps. As Figure 7(b) illustrates, this process of adding and removing layers is repeated in the succeeding cycles. The reason for the credit-based mechanism's relatively slower convergence is its use of incremental rate changes in response to network feedback. Note, however, that unlike the rate-based mechanism, the credit-based mechanism is able to achieve 100% utilization of all links.

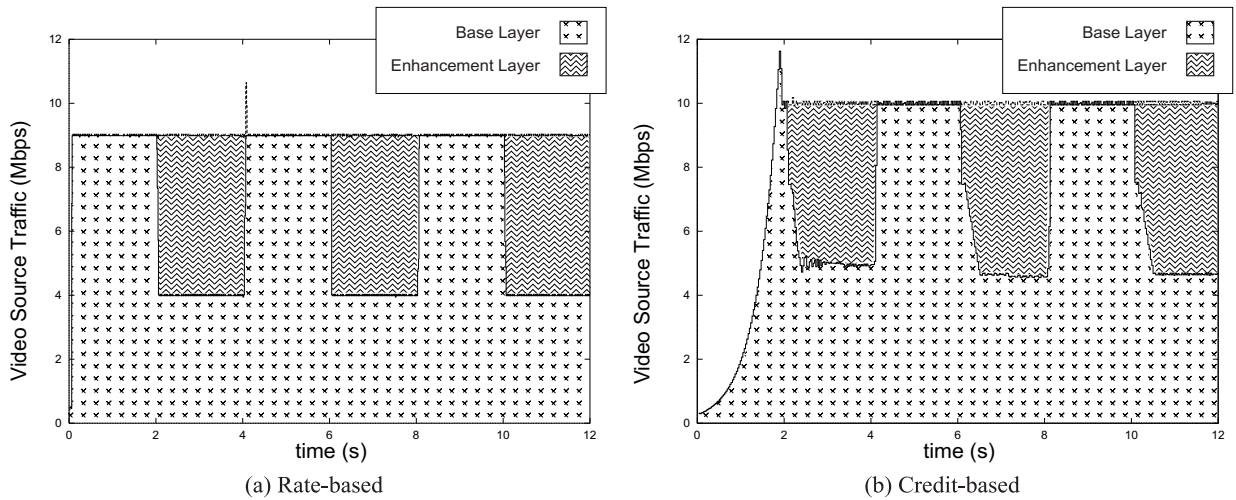

Figure 7: Creation and deletion of video layers

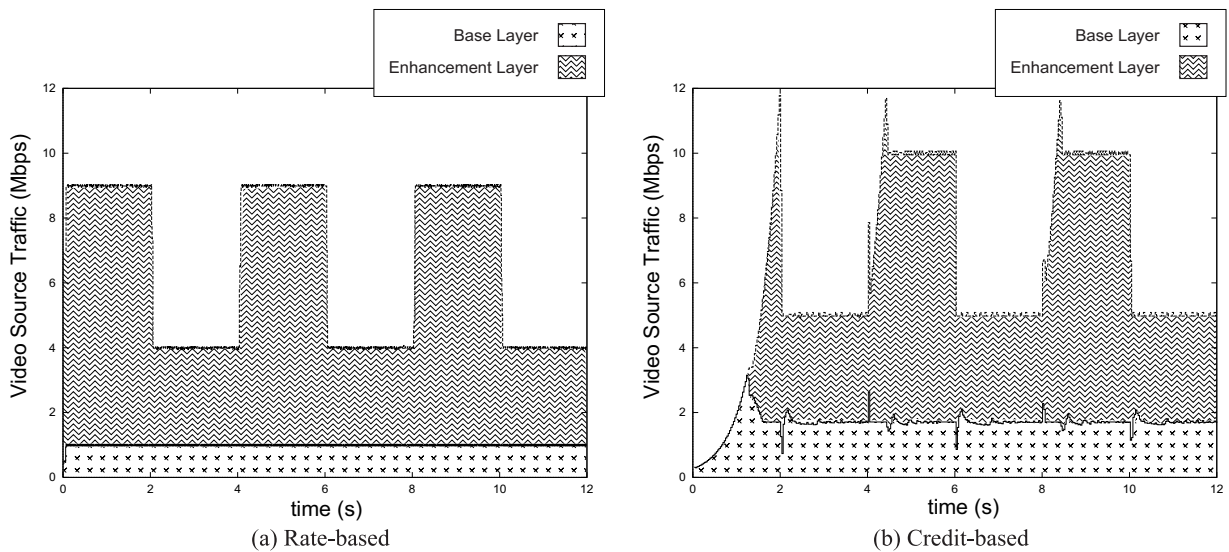

Figure 8: Inreasing and decreasing the rates of video layers

The results of this first responsiveness experiment illustrate how the rate-based and credit-based mechanisms are able to respond to changes in network congestion by adding or removing an enhancement layer of video. The rate-based mechanism, due to its use of explicit rate feedback, is able to respond more quickly than the credit-based mechanism to sudden changes in network bandwidth availability, but the credit-based mechanism is able to achieve higher link utilization.

In the second experiment, persistent interfering traffic is applied at a rate of 98 Mbps on link $L_1$. The square-wave interfering traffic applied to link $L_2$ in the first experiment is also applied in the second experiment. With 2 Mbps available on link $L_1$ and between 5 and 10 Mbps available on link $L_2$, it is expected that both feedback mechanisms will result in two layers of video being transmitted at all times, but with an oscillating rate for the enhancement layer. Figure 8 displays the transmission rates of each video layer generated by the source for the rate-based and credit-based mechanisms.

Since the available bandwidth on link $L_1$ never exceeds the available bandwidth on link $L_2$, no layers are added or deleted once the the rate of the base layer has been established. In the rate-based case, the persistent interfering traffic on link $L_1$ results in a base layer generated at 1 Mbps. The oscillating interfering traffic on link $L_2$ results in an enhancement layer of video being generated by the source, with the combined rate of both layers fluctuating in concordance with the oscillations in available bandwidth on link $L_2$. Since the rate-based scheme relies on explicit rate feedback, responses to changes in available bandwidth on link $L_2$ take approximately one round trip time (10 ms) to be reflected at the source. Again, the rate-based mechanism's target utilization of 99% prevents the source from utilizing the entire amount of bandwidth available.

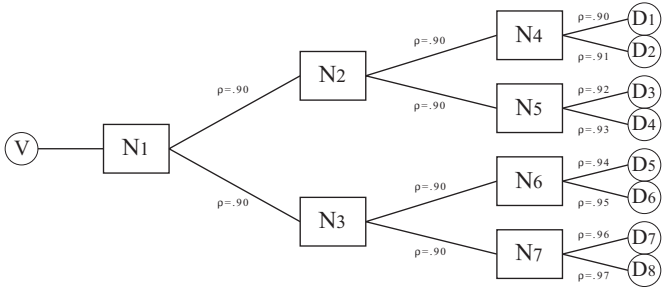

Figure 9: Simulation model for evaluating scalability.

In the credit-based case, the source starts by incrementally increasing its transmission rate and, after approximately 1.25 seconds, responds to the differential in the available bandwidths on links $L_1$ and $L_2$ by generating an enhancement layer of video. Thereafter, it takes approximately 300 ms to reduce the transmission rate of the base layer to 1.8 Mbps and takes 575 ms to increase the combined transmission rate of the base and enhancement layers up to 10 Mbps. At time $t = 2$ secs, the available bandwidth on link $L_2$ drops to 5 Mbps, and the credit-based mechanism reacts relatively quickly, bringing the combined rate of the base and enhancement layers down to 5 Mbps in approximately 40 ms. At time $t = 4$ sec, the available bandwidth on link $L_2$ returns to 10 Mbps, and the credit-based mechanism takes approximately 200 ms to converge to the new rate. The credit-based mechanism repeats this behavior in later cycles.

The results from the second experiment illustrate the ability of both feedback mechanisms to adapt the transmission rate of a layer of video when bandwidth availability in the network changes. Again, the rate-based feedback mechanism is more responsive than the credit-based mechanism due to its use of explicit rate feedback, but achieves poorer throughput due to its use of a target link utilization less than 100%.

### 4.2 Scalability

Scalability is perhaps the most important performance measure of a multicast mechanism. Two forms of scalability are studied: (1) the scalability of video quality as the number of destinations exceeds the number of video layers the network and end systems can support, and (2) the scalability of video quality as the network's end-to-end propagation delay increases.

A network model formed of a binary tree topology is used to study these two forms of scalability and is shown in Figure 9. It consists of one video source $V$, eight destinations $D_1,...,D_8$, and seven intermediate nodes $N_1,...,N_7$. Independent streams of Poisson interfering traffic with an average rate of 90 Mbps are applied on all links interconnecting intermediate nodes. Poisson interfering traffic streams are also applied to all links connecting intermediate nodes to destinations. The amount of interfering traffic on these links is increased for each destination, with the first destination receiving an interfering Poisson load of 0.90 and the eighth destination receiving a load of 0.97.

It is important to observe that in a multicast connection the potential number of destinations may be much larger than the number of priority levels a network can support or the number of layers a video source can generate. Thus, the first scalability experiment is designed to evaluate the performance of the proposed mechanisms when the number of destinations is larger than the maximum number of video layers. In this experiment, propagation delays of 50 µs are used on links interconnecting intermediate nodes, and delays of 5 µs are used on links connecting end systems to intermediate nodes.

Figure 10 plots the average "goodput ratio" for all destinations versus the maximum number of video layers transmitted. The goodput ratio is defined as the fraction of the available bandwidth used to transport uncorrupted video layers. To calculate the goodput ratio, the combined rate of video layers fully received by the destination is divided by the destination's average available bandwidth. Figure 10 reveals a significant result for the rate-based mechanism. In spite of failing to achieve 100% utilization, it achieves favorable goodput ratios between 65% and 80%, performing better than the credit-based mechanism when the maximum number of layers is small compared to the total number of destinations. The credit-based mechanism exhibits a slightly better goodput when the maximum number of layers is greater than five.

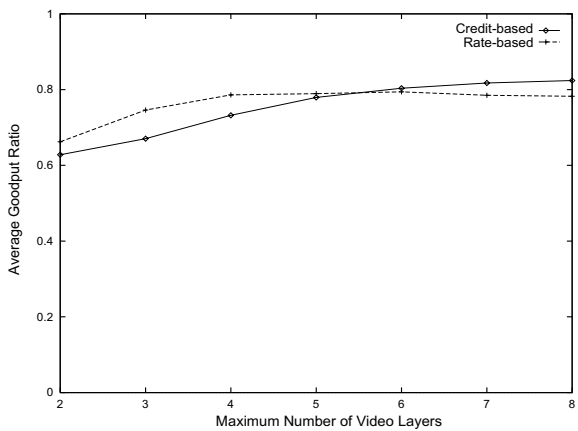

Figure 10: Average goodput ratio for all destinations vs. maximum number of layers ($L$).

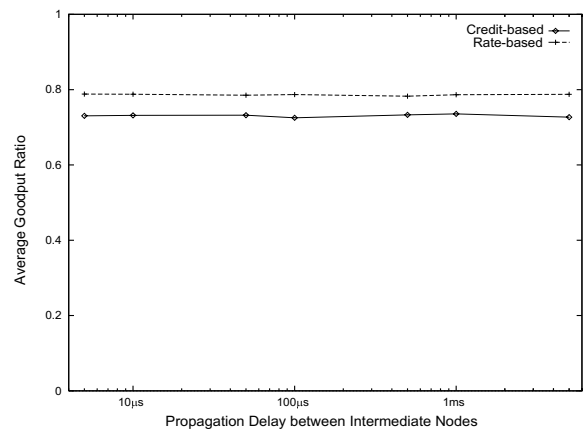

Figure 11: Average goodput ratio for all destinations vs. propagation delay between intermediate nodes ($L = 4$).

| Link $L_1, L_2$ Prop. Delay | Rate-based Mechanism | | | | Credit-based Mechanism | | | |
|---|---|---|---|---|---|---|---|---|
| | Rate | | | Fairness | Rate | | | Fairness |
| | V1 | V2 | V3 | $\sigma$ | V1 | V2 | V3 | $\sigma$ |
| 5 $\mu$s | 2.998 | 2.996 | 2.996 | 0.00135 | 3.244 | 3.244 | 3.243 | 0.00008 |
| 50 $\mu$s | 2.986 | 2.984 | 2.984 | 0.00147 | 3.243 | 3.244 | 3.244 | 0.00059 |
| 500 $\mu$s | 3.023 | 3.019 | 3.018 | 0.00250 | 3.243 | 3.243 | 3.243 | 0.00008 |
| 5 ms | 2.990 | 2.985 | 2.983 | 0.00366 | 3.242 | 3.241 | 3.242 | 0.00057 |

Table 3: Video transmission rates and fairness metric with Poisson interfering traffic.

The rate-based mechanism's average goodput ratio received by all eight destinations improves as the maximum number of layers increases from 2 to 4. After this point, the mechanism's goodput ratio remains the same. It is believed that this occurs because the rate-based mechanism's 99% target utilization and its feedback packet overhead prevent any substantial gains in overall goodput ratio once a certain value has been reached. For the credit-based mechanism, the average goodput ratio increases as the maximum number of video layers $L$ increases. This result is expected since as $L$ increases, more layers can be generated to better serve destinations with different available bandwidths.

The second scalability experiment evaluates the impact of increasing end-to-end propagation delays on the goodput ratio. The propagation delays of links interconnecting intermediate nodes are varied from 5 $\mu$s to 5 ms, and the maximum number of layers is set to 4. Figure 11 illustrates how the goodput ratio varies according to the network size. An excellent result is achieved for both mechanisms in this experiment. There is basically no change in the goodput ratio achieved in a LAN environment with propagation delays of 5 $\mu$s and in a WAN environment with propagation delays of 5 ms. We believe there are two factors contributing to the flatness of these curves. First, video traffic is continuous in nature and therefore generates a steady stream of returning feedback packets. Second, by using multi-layered video, losses — even though they increase at higher propagation delays — are isolated to the lowest priority layers of the video stream and therefore have less impact on the goodput ratio.

The scalability results presented in this section are encouraging. Both mechanisms exhibit goodput ratios of approximately 70% when a small number of layers is allowed and 80% when a larger number of layers is allowed. This signifies that most of the available bandwidth to any destination is being used to transmit uncorrupted video layers, which are most important in determining the video quality.

### 4.3 Fairness

An important factor in the evaluation of any traffic control mechanism is its fairness. If the mechanism fails to divide bandwidth equally among competing connections, then some connections may unfairly receive better service than others. This set of simulation experiments evaluates how fairly the two proposed feedback mechanisms allocate bandwidth to competing video connections.

The so-called "parking lot" model depicted in Figure 12 is used to test the fairness of the rate-based and credit-based feedback mechanisms. This network topology consists of three video sources $V_1,...,V_3$, each located at a different point in the network and transmitting video across intermediate nodes $N_1,...,N_3$ to a common destination $D$. Links $L_1$, $L_2$ and $L_3$ are congested with independent interfering Poisson traffic loads of $\rho = 0.90$. This leaves, on average, 10 Mbps of available bandwidth on each of the bottleneck links. In order to measure the effect of the round trip time on the fairness of the feedback mechanisms, propagation delays between intermediate nodes are varied between 5 $\mu$s and 5 ms, representing distances of 1 km and 1000 km, respectively.

The allocation of bandwidth to competing video traffic streams is said to be optimal if it is *max-min fair*. A max-min fair allocation of bandwidth occurs when all active connections not bottlenecked at an upstream node are allocated an equal share of the available bandwidth at every downstream node [14]. In the model shown in Figure 12, a max-min fair allocation of bandwidth occurs if all three sources are told to transmit at the same rate. To measure fairness, we calculate the standard deviation $\sigma$ of the rates that each source transmits across the bottleneck link $L_3$. An optimally fair allocation results in a standard deviation of zero.

Table 3 summarizes the results of the fairness simulations. It presents the average bit rate in Mbps used by each video traffic stream on bottleneck link $L_3$. Since the average available bandwidth on link $L_3$ is 10 Mbps, the optimal fair share is 3.333 Mbps for each of the three video streams. Both the credit-based and rate-based mechanisms are fair in the sense that they equally divide the available bandwidth of link $L_3$. However, neither mechanism achieves the optimal fair share of 3.333 Mbps due to the impact of feedback delay and, in the case of the rate-based mechanism, a 99% target utilization. Nevertheless, because each source receives nearly the same amount of bandwidth on the bottleneck link, the allocation of bandwidth is said to be fair.

While the credit-based mechanism shows slightly fairer performance, the fairness metrics remain very close to zero for all propagation delay values in both the rate-based and credit-based cases. These results demonstrate the fair behavior of both mechanisms, regardless of the distances from the competing video sources to a common bottleneck link.

### 5 Conclusion

Two multi-layered, feedback-based mechanisms for the transport of multicast video have been presented and investigated in this paper. In both mechanisms, the source uses network feedback to dynamically adjust both the number of video layers it generates and the rate at which each layer is generated. By doing so, it optimizes bandwidth utilization and the quality of video received by each destination. The two feedback mechanisms exhibit several trade-offs, however. While the rate-based mechanism provides better responsiveness and slightly better goodput when the maximum number of layers is low, the credit-based mechanism provides better utilization and slightly better fairness.

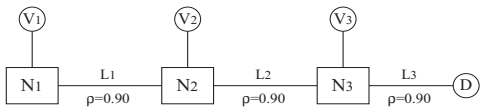

Figure 12: Simulation model for evaluating fairness

It is necessary to say a few words about applying the rate-based and credit-based mechanisms to existing networks. In ATM networks, which naturally support the notion of a virtual connection required by both mechanisms, only two priority levels are allowed. This means that without modification, ATM can only offer two layers of video per connection. Supporting more than two layers of video requires implementation of *prioritized virtual connections*. If at call admission the end system can specify a priority for a virtual connection, then end systems may establish several multicast virtual connections, one for each layer of video. The switches would then be required to preferentially discard cells from low priority connections. ATM switches must also integrate the cell merging, explicit rate calculation, and multi-layered credit passing functions required by this paper's mechanisms. Fortunately, a form of feedback packet merging has already been suggested for ABR service in ATM networks, and explicit rate feedback has been adopted for use by ABR [15]. Credit-based flow control solutions for ABR are also common in current ATM networking equipment [12].

In IP version 6 networks, eight priority levels for real-time traffic are supported. However, the types of functionalities required by the rate-based and credit-based mechanisms are largely non-existent. To solve this problem, an *active networking* solution may be applied in the Internet. Active networking, which has recently been receiving a great deal of research attention [16], allows IP routers to be dynamically programmed to perform new functions. Through the use of active networking technology, the rate-based mechanism's functions (explicit rate calculation, feedback merging) and the credit-based mechanism's functions (credit generation, reception field summation) can be implemented in IP routers. RSVP [1] may also be used in conjunction with active networking to reserve IP router resources such as bandwidth and buffer space and to ensure that the same routers traversed by feedback packets on the forward path are also traversed on the return path.

In future work, we intend to explore the impact of the mechanisms described in this paper on actual video, both through simulation and through implementation on a modified IP network testbed.